\documentclass[journal,onecolumn]{IEEEtran}
\IEEEoverridecommandlockouts
\usepackage{amsmath,amssymb,amsfonts}
\usepackage{algorithmic}
\usepackage{graphicx}
\usepackage{dblfloatfix}
\usepackage{textcomp}
\usepackage{xcolor}
\usepackage{float}
\usepackage{multirow}
\usepackage{footnote}

\usepackage{csquotes}
\usepackage{setspace}
\usepackage{subcaption}

\usepackage{booktabs}

\usepackage{acronym}

\newacro{5g}[5G]{Fifth Generation}
\newacro{6g}[6G]{Sixth Generation}
\newacro{as}[AS]{Authentication Server}
\newacro{atm}[ATM]{Automated Teller Machine}
\newacro{ai}[AI]{Artificial Intelligence}
\newacroplural{ais}[AIs]{Artificial Intelligences}
\newacro{arima}[ARIMA]{Autoregressive Integrated Moving Average}
\newacro{b5g}[B5G]{Beyond 5G}
\newacro{ban}[BAN]{Body Area Network}
\newacro{bsi}[\textit{BSI}]{\textit{Federal Office for Information Security}}
\newacro{bdr}[BDR]{Bit Disagreement Rate}
\newacro{bs}[BS]{Base Station}
\newacro{ber}[BER]{Bit Error Rate}
\newacro{bsr}[BSR]{Bit Success Rate}
\newacro{ca}[CA]{Certification Authority}
\newacro{cav}[CAV]{Connected Autonomous Vehicles}
\newacro{cc}[CC]{Common Criteria}
\newacro{cir}[CIR]{Channel Impulse Response}
\newacro{cr}[CR]{Challenge-Response}
\newacro{cpu}[CPU]{Central Processing Unit}
\newacro{cpps}[CPPS]{Cyber-Physical Production System}
\newacro{crl}[CRL]{Certificate Revocation List}
\newacro{csi}[CSI]{Channel State Information}
\newacro{crke}[CRKE]{Channel-Reciprocity Based Key Extraction}
\newacro{ctf}[CTF]{Channel Transfer Function}
\newacro{cotf}[COTF]{Commercial-off-the-Shelf}
\newacro{cmos}[CMOS]{Complementary Metal-Oxide-Semiconductors}
\newacro{dos}[DoS]{Denial-of-Service}
\newacro{ddos}[DDoS]{Distributed-Denial-of-Service}
\newacro{dna}[DNA]{Deoxyribonucleic Acid}
\newacro{dtls}[DTLS]{Datagram Transport Layer Security}
\newacro{dct}[DCT]{Discrete Cosine Transformation}
\newacro{dlt}[DLT]{Distributed Ledger Technology}
\newacro{dwt}[DWT]{Discrete Wavelet Transform}
\newacro{eal}[EAL]{Evaluation Assurance Level}
\newacro{ecc}[ECC]{Elliptic Curve Cryptography}
\newacro{ecg}[ECG]{Electrocardiogram}
\newacro{eeg}[EEG]{Electroencephalogram}
\newacro{embb}[eMBB]{enhanced Mobile broad-Band}
\newacro{emg}[EMG]{Electromyogram}
\newacro{eog}[EOG]{Electrooculography}
\newacro{enb}[eNodeB]{Evolved Node B}
\newacro{er}[ER]{Extended Reality}
\newacro{fpga}[FPGA]{Field Programmable Gate Array}
\newacro{fdd}[FDD]{Frequency Division Duplexing}
\newacro{fp}[FP]{true positive}
\newacro{fn}[FN]{true negative}
\newacro{gdpr}[GDPR]{General Data Protection Regulation}
\newacro{gd}[G\&D]{Giesecke \& Devrient}
\newacro{h2m}[H2M]{Human-to-Machine}
\newacro{h2s}[H2S]{Human-to-Service}
\newacro{hmac}[HMAC]{Keyed-Hash Message Authentication Code}
\newacro{htc}[HTC]{Hologaphic-Type Communication}
\newacro{hotp}[HOTP]{HMAC-based One-time Password Algorithm}
\newacro{hsm}[HSM]{Hardware Security Module}
\newacro{ics}[ICS]{Industrial Control System}
\newacro{iacs}[IACS]{Industrial Automation and Control System}
\newacro{ioe}[IoE]{Internet of Everything}
\newacro{iiot}[IIoT]{Industrial Internet of Things}
\newacro{iot}[IoT]{Internet of Things}
\newacro{iobnt}[IoBNT]{Internet of Bio-Nano Things}
\newacro{io}[I/O]{Input/Output}
\newacro{ic}[IC]{Integrated Circuit}
\newacro{id}[ID]{Identificator}
\newacro{ids}[IDS]{Intursion Detection System}
\newacro{irs}[IRS]{Intelligent Reflecting Surface}
\newacro{istn}[ISTN]{Integrated Space and Terrestrial Network}
\newacro{it}[IT]{Information Technology}
\newacro{itu}[ITU]{International Telecommunication Union}
\newacro{jcop}[JCOP]{Java Card Open Platform}
\newacro{kba}[KBA]{Knowledge Based Authentication}
\newacro{kdf}[KDF]{Key Derivation Function}
\newacro{kf}[KF]{Kalman Filter}
\newacro{led}[LED]{Light Emitting  Diode}
\newacro{lte}[LTE]{Long Term Evolution}
\newacro{ltea}[LTE-A]{Long Term Evolution Advanced}
\newacro{lr}[LR]{Linear Regression}
\newacro{los}[LoS]{Line of Sight}
\newacro{lorawan}[LoRaWAN]{Long Range Wide Area Network}
\newacro{mbb}[MBB]{Mobile Broadband}
\newacro{mc}[MC]{Molecular Communication}
\newacro{mfa}[MFA]{Multi-Factor Authentication}
\newacro{mcc}[MCC]{Mobile Cloud Computing}
\newacroplural{mcus}[MCUs]{Microcontroler Units}
\newacro{m2m}[M2M]{Machine-to-Machine}
\newacro{m2s}[M2S]{Machine-to-Service}
\newacro{mimo}[MIMO]{Multiple Input Multiple Output}
\newacro{mmimo}[mMIMO]{massive Multiple Input Multiple Output}
\newacro{ml}[ML]{Machine Learning}
\newacro{mulc}[mULC]{massive Ultra-Reliable Low-Latency Communication}
\newacro{mmtc}[MMTC]{massive Machine Type Communication}
\newacro{mmg}[MMG]{Mechanomyogram}
\newacro{multos}[MULTOS]{Multii-Application Smart Card Operating System}
\newacro{mux}[MUX]{Multiplexer}
\newacro{mnc}[MNC]{Mobile Network Code}
\newacro{me}[ME]{Mobile Environment}
\newacro{mac}[MACs]{Message Authentication Codes}
\newacro{mps}[MPS]{Master Production Schedule}
\newacro{ngmn}[NGMN]{Next Generation Mobile Network}
\newacro{nic}[NIC]{Network Interface Controller}
\newacro{nist}[NIST]{National Institute of Standards and Technology}
\newacro{oath}[OATH]{Open Authentication}
\newacro{ocra}[OCRA]{\ac{oath} Challenge-Response Algorithm}
\newacro{ocsp}[OCSP]{Online Certificate Status Protocol}
\newacro{otp}[OTP]{One-Time Password}
\newacro{ook}[OOK]{On Off Keying}

\newacro{pvc}[PVC]{Polyvinyl chloride}
\newacro{pa}[PA]{Process Automation}
\newacro{pap}[PAP]{Password-Authentication-Protocol}
\newacro{physec}[PhySec]{Physical Layer Security}
\newacro{pfs}[PFS]{Perfect Forward Secrecy}
\newacro{pin}[PIN]{Personal Identification Number}
\newacro{pkc}[PKC]{Public Key Cryptography}
\newacro{pki}[PKI]{Public Key Infrastructure}
\newacro{ppg}[PPG]{Photoplethysmography}
\newacro{prng}[PRNG]{Pseudo Random Number Generator}
\newacro{puf}[PUF]{Physically Unclonable Function}
\newacroplural{pufs}[PUFs]{Physically Unclonable Functions}
\newacro{pla}[PLA]{Physical Layer Authentication}
\newacro{plc}[PLC]{Programmable Logic Controller}
\newacro{per}[PER]{Packet Error Rate}
\newacro{psr}[PER]{Packet Success Rate}

\newacro{qr}[QR]{Quick Response}
\newacro{rat}[RAT]{Radio Access Technology}
\newacro{radius}[RADIUS]{Remote Authentication Dial-In User Service}
\newacro{ram}[RAM]{Random-Access Memory}
\newacro{ran}[RAN]{Radio Access Networks}
\newacro{rf}[RF]{Radio-Frequency}
\newacro{rfid}[RFID]{Radio-Frequency Identification}
\newacro{ris}[RIS]{Reconfigurable Intelligent Surface}
\newacro{rng}[RNG]{Random Number Generator}
\newacro{ro}[RO]{Ring-Oscillator}
\newacro{rom}[ROM]{Read-Only Memory}
\newacro{rs}[RS]{Reed-Solomon}
\newacro{rsa}[RSA]{Rivest-Shamir-Adleman}
\newacro{rssi}[RSSI]{Received Signal Strength Indicator}
\newacro{rsrp}[RSRP]{Reference Signal Received Power}
\newacro{re}[RE]{Resource Elements}

\newacro{maf}[MAF]{Moving Average Filter}
\newacro{snr}[SNR]{Signal-to-Noise Ratio}
\newacro{sdn}[SDN]{Software-Defined Network}
\newacro{sdr}[SDR]{Software-Defined Radio}
\newacro{seccos}[SECCOS]{Secure Chip Card Operating System}
\newacro{sip}[SIP]{Session Initiation Protocol}
\newacro{skg}[SKG]{Secret Key Generation}
\newacro{sram}[SRAM]{Static Random Access Memory}
\newacro{srs}[SRS]{Software Radio Systems}
\newacro{starcos}[STARCOS]{Smart Card Chip Operating System}
\newacro{sha}[SHA]{Secure Hash Algorithm}
\newacro{se}[SE]{Static Environment}
\newacro{svm}[SVM]{Support Vector Machine}
\newacro{tcg}[TCG]{Trusted Computing Group}
\newacro{tpm}[TPM]{Trusted Platform Module}
\newacro{tls}[TLS]{Transport Layer Security}
\newacro{trng}[TRNG]{True Random Number Generator}
\newacro{tsn}[TSN]{Time-Sensitve Networking}
\newacro{tofu}[TOFU]{Trust On First Use}
\newacro{tufu}[TUFU]{Trust Upon First Use}
\newacro{totp}[TOTP]{Time-based One-time Password Algorithm}
\newacro{tia}[TIA]{Totally Integrated Automation}
\newacro{tp}[TP]{true positive}
\newacro{tn}[TN]{true negative}
\newacro{uav}[UAV]{Unmanned Arial Vehicles}
\newacro{usb}[USB]{Universal Serial Bus}
\newacro{usrp}[USRP]{Universal Software Radio Peripheral}
\newacro{uhd}[UHD]{USRP Hardware Driver}
\newacro{usim}[USIM]{Universal Subscriber Identity Module}
\newacro{ue}[UE]{User Equipment}
\newacro{urllc}[URLLC]{Ultra-Reliable Low-Latency Communication}
\newacro{ulbc}[ULBC]{Ultra-Reliable Low-Latency Broadband Communication}
\newacro{umbb}[uMBB]{ubiquious Mobile Broadband}
\newacro{ummimo}[UM-MIMO]{Ultra-Massive MIMO}
\newacro{vlc}[VLC]{Visible Light Communication}
\newacro{warp}[WARP]{Wireless open-Access Research Platform}
\newacro{wt}[WT]{Wavelet Transform}

\usepackage[
	pdfborder={0 0 0},
	colorlinks=false,%
	urlcolor=black,%
	linkcolor=black,%
	citecolor=black,%
	filecolor=black,%
	breaklinks,%
	]{hyperref}%
	
\usepackage{cite} 
\usepackage{cite}  

\def\BibTeX{{\rm B\kern-.05em{\sc i\kern-.025em b}\kern-.08em T\kern-.1667em\lower.7ex\hbox{E}\kern-.125emX}}
    
\usepackage{indentfirst}

\begin{document} 

\title{
Experimental Analysis of Microbubble Propagation for In-Body Data Transmission





}

\author{
\IEEEauthorblockN{Annika Tjabben\IEEEauthorrefmark{1}, Lea Bergkemper\IEEEauthorrefmark{3}, Carolin Conrad\IEEEauthorrefmark{1}, Michael Gundall\IEEEauthorrefmark{1}, and Hans D. Schotten\IEEEauthorrefmark{1}\IEEEauthorrefmark{2}} \\
\IEEEauthorblockA{\IEEEauthorrefmark{1}German Research Center for Artificial Intelligence (DFKI), Germany\\ E-mail: \{Annika.Tjabben, Carolin.Conrad, Michael.Gundall, Hans.Schotten\}@dfki.de} \\
\IEEEauthorblockA{\IEEEauthorrefmark{3} 
E-mail: \{bergkemper.l25\}@gmail.com} \\
\IEEEauthorblockA{\IEEEauthorrefmark{2}RPTU University Kaiserslautern-Landau, Germany\\
E-mail: \{Schotten\}@rptu.uni-kl.de} 

}

\maketitle

%
%
%
%
\begin{abstract}
In-body communication is an upcoming field with significant implications for medical diagnostics and therapeutic interventions. Microbubbles have gained attention due to their distinct physical properties, making them promising candidates to facilitate communication within the human body. 
This work explores the use of microbubbles as communication carriers, with a particular focus on their detection and the application of a modulation scheme. Through experimental analysis the feasibility and effectiveness of microbubble-based communication is tested. Filtering and peak detection methods are applied to accurately identify the presence of microbubbles despite noise, demonstrating the feasibility of microbubble-based communication systems for future biomedical applications. The results offer insights into signal integrity, noise challenges, and the optimization of detection algorithms, providing a foundation for future advancements in this field. 
\end{abstract}



%
%
%
%

\section{Introduction} 
\label{sec:introduction}

\noindent As healthcare increasingly shifts toward personalized and precision medicine, the ability to monitor and communicate using in-body systems gains importance. \ac{mc} offers an approach to facilitate this, mimicking biological processes that already exist within the human body. The \ac{iobnt} framework leverages this capability by integrating nanotechnology, biotechnology, and communication systems to enable a new era of smart medical devices and therapies, ranging from continuous health monitoring to controlled drug release. Hence, ongoing research focuses on investigating novel methods of data exchange. 
This work explores the use of microbubbles as a communication medium within biological environments; however, the initiation of communication still relies on external stimuli. 

Microbubbles are tiny gas-filled spheres encapsulated by a shell of lipids, proteins, or polymers, and are already well-established in clinical practice due to their high echogenicity. Currently they are primarily used as contrast agents in ultrasound imaging, enabling early disease detection and supporting therapeutic applications such as targeted drug or gene delivery~\cite{Lee2017}. 
The advantageous properties of the bubbles include biocompatibility, a small size of approximately 2.5\,\textnormal{\textmu{}m}, and natural degradation within the body. These characteristics make them a safe and versatile tool in medical applications~\cite{Frinking2020}. 
Given their unique features, microbubbles present an ideal candidate for in-body communication. Their ability to be tracked using non-invasive imaging techniques allows for monitoring, while their capability to encapsulate gases or therapeutic agents adds a functional dimension. Therefore, microbubbles are potential dual-role agents, capable of both carrying therapeutic payloads and facilitating molecular data exchange. 

While previous studies have mainly focused on their diagnostic and therapeutic applications, this research evaluates the feasibility of using microbubbles as communication agents by observing their behavior and tracking mechanisms on an experimental basis~\cite{Lee2017}. 
The transmission of data is enabled by means of the injection of microbubbles, which represent a ``high'' signal. Conversely, the absence of particles indicates a ``low'' signal. The present work utilizes an ultrasound approach for the detection of signals and their subsequent use for \ac{mc}. The primary aim is to establish the foundations for future advancements in the field, as the overall findings have implications for the development of smart, biocompatible communication systems integrated with diagnostic and therapeutic functionalities. 


This work is organized as follows: \autoref{sec:state} provides an overview of the field of \ac{mc} and describes the relevant research in this area. Moreover, \autoref{sec:setup} presents the experimental configuration for microbubble detection and details measurements and encountered issues. In \autoref{sec:results}, two different approaches to signal detection are presented, along with the results obtained from implementing each. 
In \autoref{sec:coding} analyzes system performance and coding implications.
Finally, in \autoref{sec:conlusion}, conclusions of this work are outlined, and suggestions for future research are proposed.

\section{Related Work} 
\label{sec:state}

\noindent Current research on microbubbles focuses primarily on ultrasound-based detection for medical imaging and therapy. 
Microbubbles oscillate under transmitted ultrasound pulses, generating detectable acoustic responses that enable high sensitivity and non-invasive tracking. Recent advances have improved imaging resolution and depth, supporting targeted drug delivery and molecular diagnostics~\cite{brown2019investigation,10124036}. 
Beyond imaging, ultrasound enables precise control of microbubbles, including trapping, self-assembly, and navigation of microrobot swarms against blood flow in mouse brain vasculature~\cite{DelCampoFonseca2023}, as well as cyclic jetting for low-pressure intracellular drug delivery~\cite{Cattaneo2025}. 

The concept of the \ac{iobnt}, where engineered biological cells communicate through molecular signals to form functional networks for applications such as health monitoring and environmental sensing, was first introduced in foundational work on molecular communication systems~\cite{Akyildiz2015}. 

While advancements in microbubble detection for imaging and therapy abound \cite{brown2019investigation,10124036,Wang2019}, their adaptation as information carriers in \ac{mc} remains limited.
Recent surveys highlight \ac{mc} channel modelling for diffusive particles \cite{farsad2016comprehensive}, but experimental validation with trackable agents like microbubbles is limited. Emerging work explores bio-cyber interfaces in \ac{iobnt} \cite{Akyildiz2015,Hofmann2025BioCyberInterface}, yet lacks modulation and detection schemes for ultrasound-detectable carriers. This gap motivates the present experimental investigation of \ac{ook} modulation with a filtering-based detection. 

Beyond conventional ultrasound imaging, other detection approaches have emerged, including acoustic and visual techniques. Acoustic detection involves analyzing the unique frequency signatures generated by microbubble oscillations, enabling live tracking and characterization~\cite{1588397,cleveland2012physics}. Visual methods, such as high-speed microscopy and laser scattering, offer detailed insights into microbubble dynamics like size distribution, motion, and collapse behavior, which are critical for optimizing their performance in various applications~\cite{izadifar2019ultrasound,chen2011ultra}. However, these methods are generally more complex and less commonly used compared to ultrasound-based techniques, primarily due to practical limitations in clinical settings.


While these methods for detecting and characterizing microbubbles already exist, their focus remains predominantly on medical diagnostics and drug delivery applications. Few research papers address microbubbles as a communication medium in the context of \ac{mc}. 

\section{Testbed and Measurements} 
\label{sec:setup}

\noindent To evaluate the feasibility of microbubbles as information carriers, a reliable and robust experimental testbed is required. Therefore, a measurement framework is developed, which is designed to provide a controlled environment, enabling precise analysis under various conditions. 

\subsection{Experimental Testbed for Data Transmission} 

\begin{figure} [htb]
    \centering
    \includegraphics[width=0.8\linewidth]{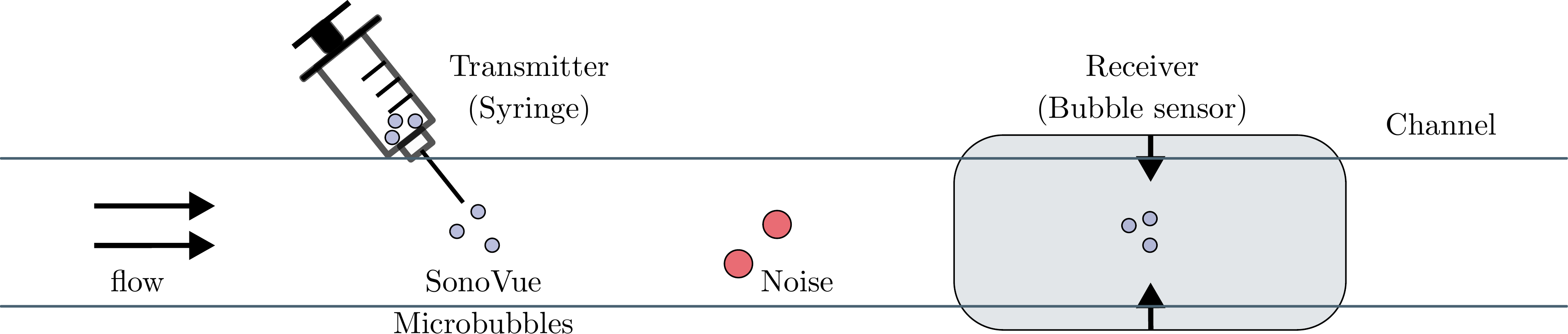} 
    \caption{Schematic representation of data transmission via microbubbles.} 
    \label{img:channel}
\end{figure}

\begin{figure}[b] 
\centering
    \includegraphics[width=0.5\linewidth]{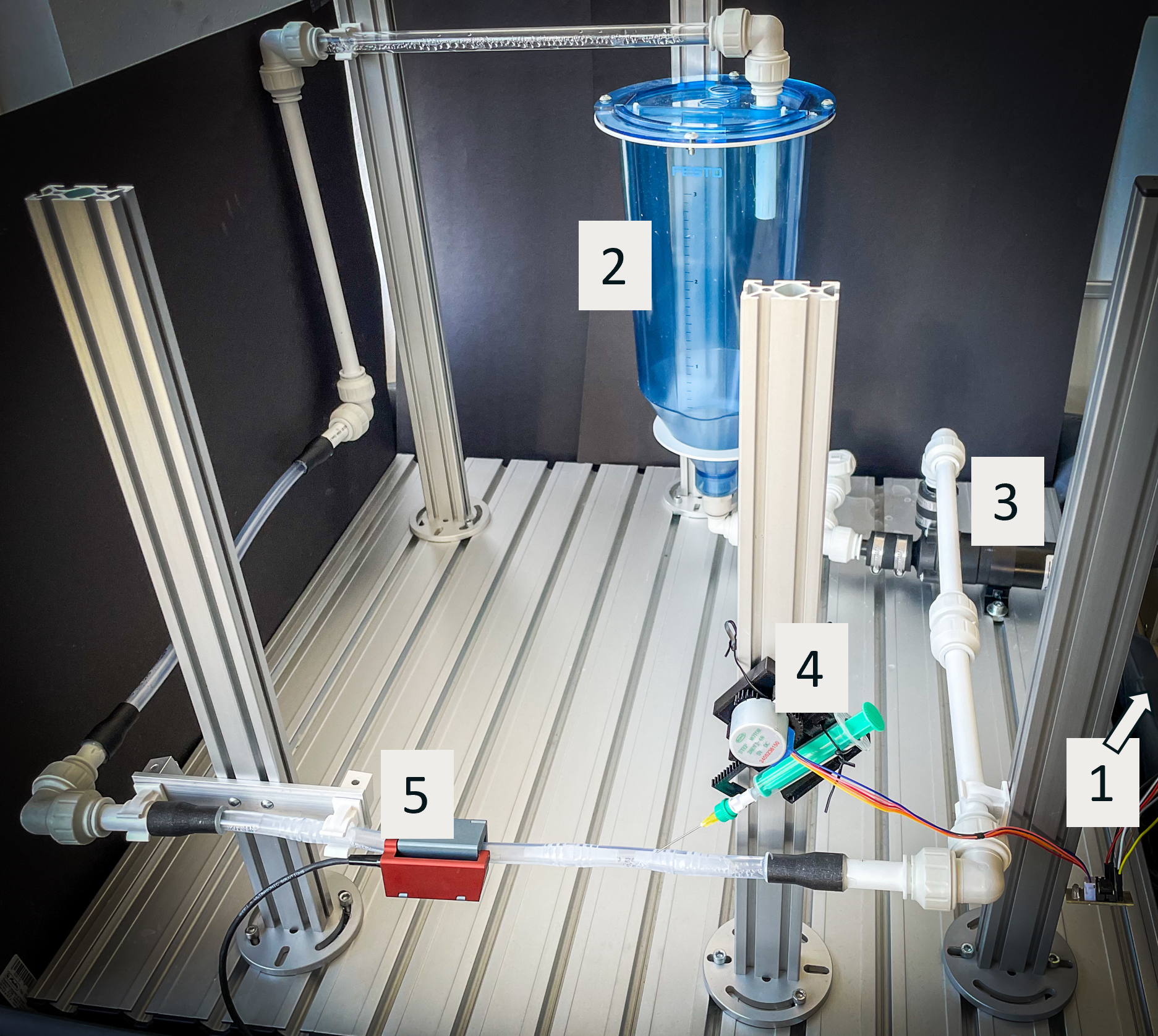} 
\caption{Setup for Microbubble Injection and Detection \\ (1: Siemens S7-1500 PLC, 2: Water Tank, 3: Water Pump, 4:~Syringe for Microbubble Injection, 5: Bubble Sensor).} 
\label{img:expsetup}
\end{figure}

The fundamental structure of the system comprises a transmission channel that utilizes microbubbles as signal molecules, a transmitter which is represented by syringes for injecting the signal molecules and a receiver for detecting the transmitted data. Fig.~\ref{img:channel} provides a schematic overview of this system, serving as a conceptual foundation for the experimental setup depicted in Fig.~\ref{img:expsetup}. 

To align the system with a real application the tubes employ a continuous loop, thereby allowing microbubbles to circulate. The tubes have a diameter of 3/8\,inch. Although this is larger than typical human blood vessels, the bigger size enhances visualization and closely mimics the dynamic flow conditions of biological systems. Moreover, key dimensionless parameters such as the Reynolds and Péclet numbers remain within physiologically relevant ranges, ensuring that the observed transport behavior of microbubbles is scalable and representative of in-body conditions~\cite{farsad2016comprehensive}. 

To precisely control and monitor the system a Festo Master Production Schedule Process Automation Compact-Workstation with a Siemens S7-1500 Programmable Logic Controller is used. Furthermore, the SPX Flow CM30P7-1 pump (see Fig.~\ref{img:expsetup}) ensures a stable flow rate.
Operating within a control range of 0\,V to 10\,V, the pump is set to 3\,V for the measurements, achieving an approximate flow rate of 1.24\,L/min. 
This flow rate is low compared to typical human blood flow (5–6\,L/min), however it is inevitable to reduce the flow rate due to the higher scaled tubes and the requirements of the ultrasonic sensor used for tubes this size~\cite{Rallabandi2015Streaming}. 
The SonoVue microbubbles used as a signaling molecule for the measurements are a widely utilized medical contrast agent developed by BRACCO for ultrasound imaging. 
They are created by mixing a powder containing colfosceril stearate and a palmitic acid with a sodium chloride solution, resulting in sulfur hexafluoride-filled microbubbles~\cite{sonovue2021}. 
Slightly soluble in water, these microbubbles are highly detectable via ultrasound due to their ability to reflect sound waves at the water interface. Each milliliter of solution contains approximately 8\,\textnormal{\textmu{}L} of microbubbles, with an average diameter of 2.5\,\textnormal{\textmu{}m} per microbubble~\cite{sonovue2021}. 

To create a ``high'' signal, microbubbles are injected into the system via an Arduino-controlled motorized syringe, enabling a precise and consistent injection. 
The detection is done using the GAMPT BubbleCounter BCF300, a high-precision, non-invasive Doppler ultrasound sensor designed for air bubble analysis in extracorporeal circulation~\cite{gampt2024}. 
For the experiment, the sensor is mounted on a PVC tube downstream of the injection point to track microbubbles. 
The system processes the reflected sound waves by summing their values and calculating an envelope function. 
Data is recorded over successive 40\,ms intervals, enabling characterization of both bubble size and their dynamic behavior under acoustic excitation.

\subsection{Framework Conditions and Limitations} 

To ensure consistent and controlled measurements, the parameters of the setup are optimized to the use case of data transmission. 
Distilled water is employed, as it ensures that microbubbles decay within the time specified by the manufacturer, preventing system flooding, defined here as the excessive buildup of residual bubbles that saturates the channel and degrades measurement accuracy. To minimize the impact of noise due to air bubbles, the water level of the tank is maintained at a minimum height and the pump speed is restricted, allowing sufficient water movement to observe microbubble while minimizing turbulences. To achieve a stable microbubble concentration, 90\,nL of microbubble solution is diluted in 5\,mL of distilled water before injection. 

\begin{figure*}[bt]
\centering
    \begin{subfigure}[h]{0.45\textwidth} 
    \includegraphics[width=1\textwidth]{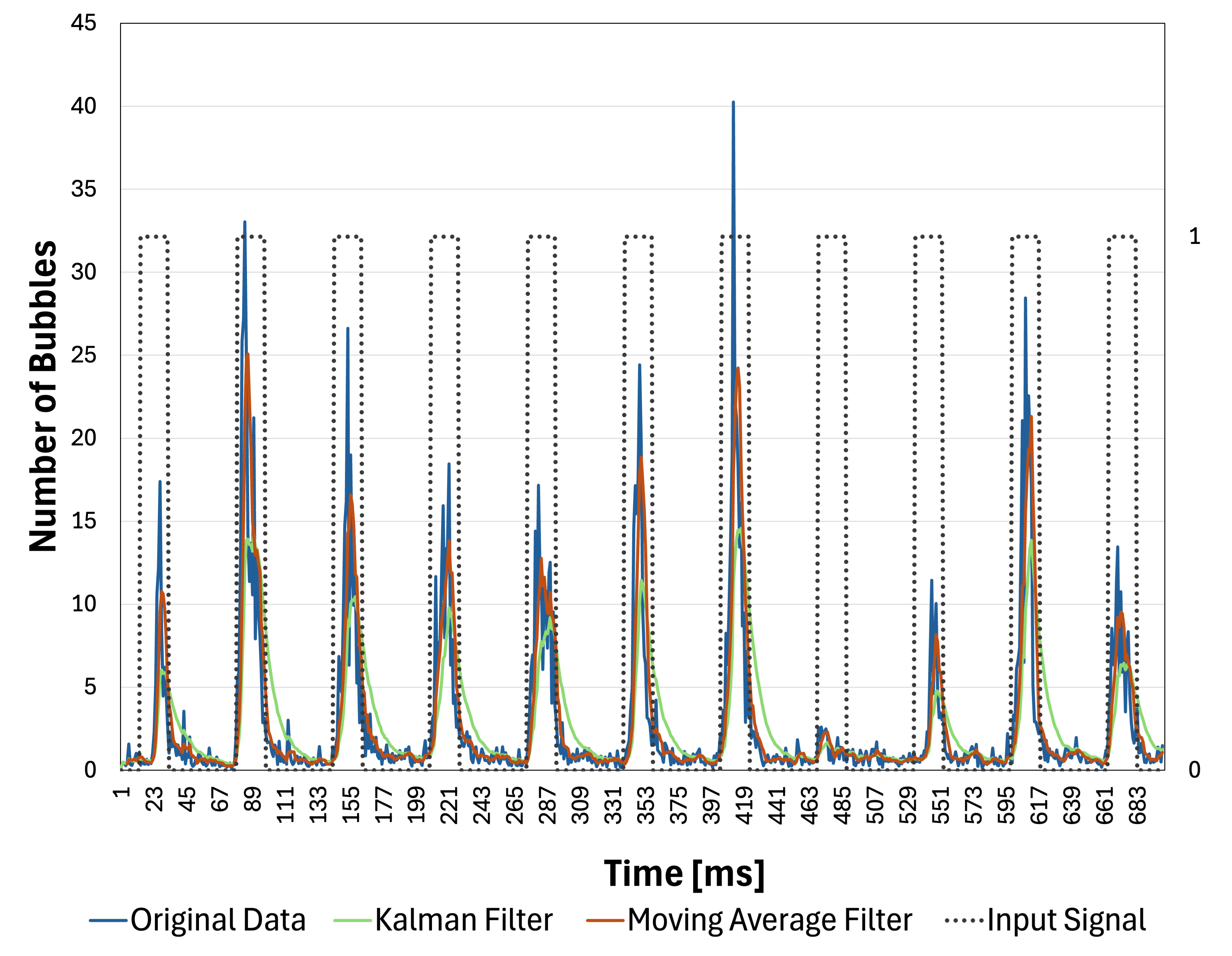}
    \subcaption{Raw signal (blue) shows periodic peaks corresponding to 0.3~s ``on'' injections separated by 2.0~s ``off'' phases, overlaid with ground truth square-wave input (black). Both moving-average (red) and Kalman filtering (green) effectively smooth environmental noise and injection variability while preserving peak structure.} 
    \label{img:filters}
    \end{subfigure}
    \hfill
    \begin{subfigure}[h]{0.45\textwidth} 
    \includegraphics[width=1\textwidth]{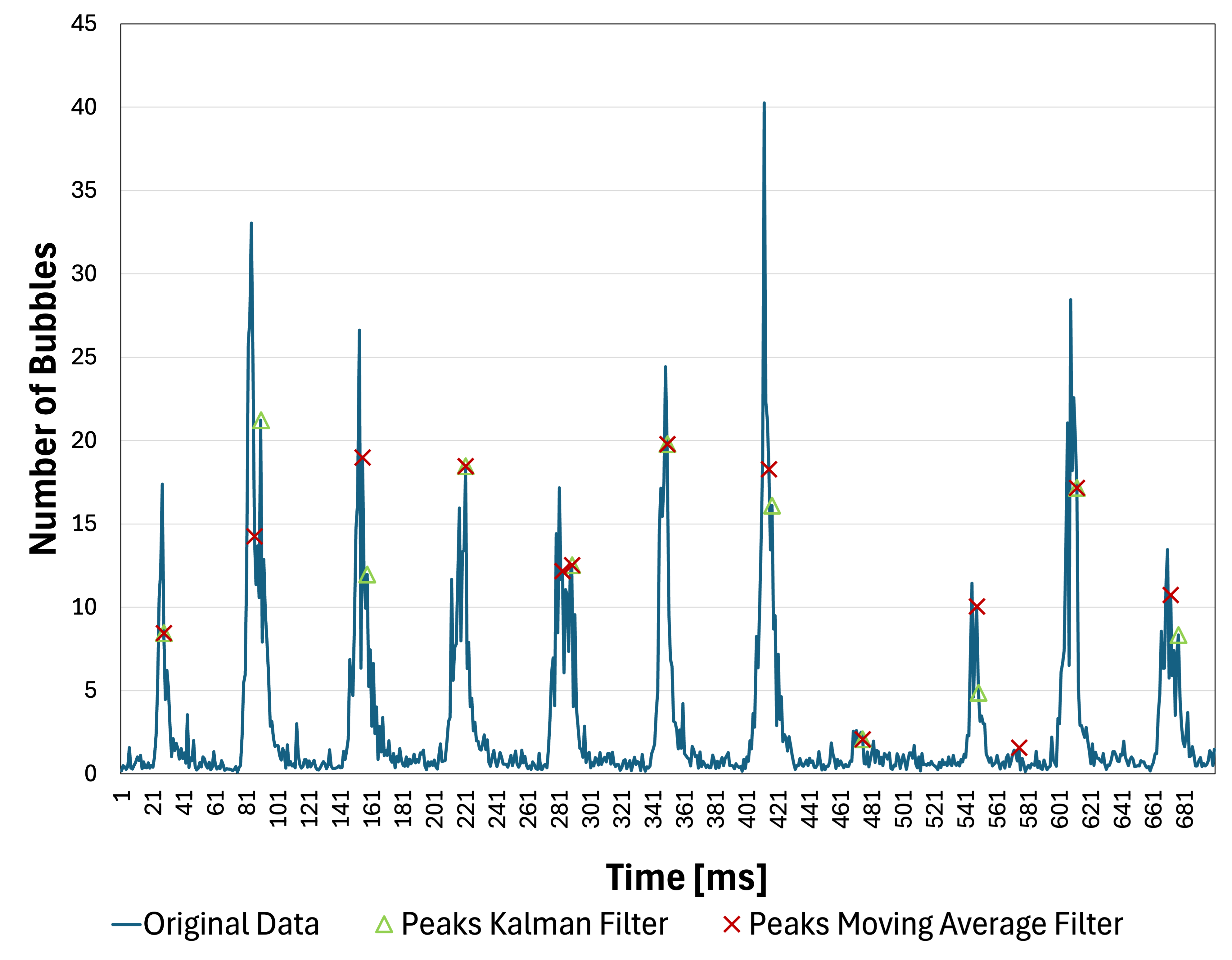}
    \subcaption{Peak Detection after applying different Filtering Techniques. The MAF (red) identifies all expected peaks but includes minor false positives from residual noise (red crosses), while the KF (green) correctly detects all peaks without false positives (green triangles).}
    \label{img:filterspeaks}
    \end{subfigure}
\caption{Captured Utrasound Data for a Periodic Injection of Microbubbles.}
\label{img:fullfilters}
\end{figure*}

Despite optimizing the system, the current setup has several limitations:
The detection of single microbubbles is challenging due to their small size, which makes it difficult to distinguish them from background noise, such as particulate matter. In contrast, air bubbles can be more easily differentiated from microbubbles because of their larger size. 
Furthermore, due to the currently used method for microbubble injection it is only possible to define the exact amount of mixture injected, while counting the number of bubbles introduced per injection is not possible, leading to signal variability and complicating reproducibility. 
The BubbleCounter module of the GAMPT system detects and counts bubbles passing a single measurement point within a 40\,ms time window~\cite{gampt2024}. As a result, differentiating microbubbles from background noise can be challenging. 
This restriction also limits the potential bit rate to 8.33\,bit/s, since at least three intervals of 40\,ms are required to reliably send an information. 







\section{Data Filtering and Performance Evaluation} 
\label{sec:results}

\noindent While the ultrasound device enables detection of microbubble clusters, the raw measurements require careful preprocessing to mitigate noise and variability before meaningful analysis can be performed. 
This is particularly challenging in dynamic environments, for instance during the injection of microbubbles into biological systems, where signal amplitudes can fluctuate due to variations in concentration and motion. Furthermore, microbubbles decay over time, leading to potential recirculation effects that interfere with newly detected values, as described in~\cite{Tjabben2024}. These residual microbubbles can distort the signal by elevating the baseline and causing disturbances. To address these challenges, preprocessing the raw time-series data, by applying filtering and smoothing techniques, is essential. These methods improve the isolation of ``high'' and ``low'' signals based on microbubble activity, and ensure accurate interpretation. 

Hence, two different filtering techniques, the \ac{maf} and the \ac{kf}, are applied for signal preprocessing. While both approaches are well-known algorithms, each implements unique strengths when applied to ultrasound time-series data with a view to minimizing the impact of noise and outliers.
The \ac{maf} (red, in Fig.~\ref{img:filters}) is one of the simplest and most widely used filtering techniques in signal processing. It works by applying a sliding window across the data to smooth out rapid fluctuations and highlight slower, underlying trends. This window shifts point by point, and for each position, the average of the data points within the window is computed, effectively reducing the noise in the signal~\cite{al2019application}. 

In comparison, the \ac{kf} (green, in Fig.~\ref{img:filters}) is a more sophisticated smoothing approach. It not only smooths data but also learns how the values evolve over time to predict future values~\cite{maybeck1990kalman}. This makes it particularly powerful in dynamic systems, where the relationship between past and present observations changes continuously.

One of the key strengths of the \ac{kf} is its adaptability through the tuning of its noise parameters $Q$, which represents the process noise (uncertainty in the model’s prediction) and $R$, which represents the measurement noise (uncertainty in the observations). A larger $Q$ allows the filter to respond more quickly to changes in the signal, reflecting greater uncertainty in the system dynamics. Conversely, a smaller $R$ places more weight on the measurements, assuming they are relatively reliable. By balancing these two parameters, the \ac{kf} can robustly track and estimate signal variations in real time. In the context of this study, the \ac{kf} can adapt to the presence of residual microbubbles in the channel, effectively minimizing the impact of prior injections and enhancing the accuracy of signal interpretation~\cite{maybeck1990kalman}.



Once the signal has been filtered to remove noise, peak detection techniques are applied to identify significant features in the data. 
SciPy’s find\_peaks() offers an advanced approach by detecting local maxima and allowing fine-tuning with parameters like height, prominence, distance, and width. It’s more robust to noise and allows for better detection of peaks with varying sizes and prominence. It considers the local signal context, making it more flexible and effective in noisy environments. However, this approach is slightly more complex to configure and requires more computation than a simple threshold.

To conduct ultrasound data and evaluate these approaches, the setup described in \autoref{sec:setup} is used. The input signal is given in form of a square-wave pulse, characterized by an ``on'' phase of 0.3\,s and an ``off'' phase of 2\,s. During the ``on'' phase, the syringe injects microbubbles into the system, resulting in a symbol duration of 2.3\,s. Apart from the limitations introduced by the ultrasound, the current setup and environmental conditions cause further restrictions. Hence, this is the shortest achievable data rate of 0.43\,bit/s. Further reducing the time between symbols leads to overlapping peaks, making it challenging to reliably distinguish between them.

Fig.~\ref{img:filters} illustrates the ultrasound data acquired during this periodic microbubble injections and the corresponding ground truth of ``high'' and ``low'' signal in form of the input signal. The data clearly show the periodic peaks corresponding to microbubble injections. However, the aforementioned challenges also become obvious in the data. For instance at 480\,ms, the signal variability becomes obvious, as one expected peak is only subtle. This might be caused by the injection method, which may not add the precise quantity of bubbles in each injection, as well as by unpredictable conditions in the fluid environment. Furthermore, the raw data contain environmental noise leading to jitter, which emphasizes the need for smoothing prior to peak detection. After applying the \ac{maf} and \ac{kf}, noise levels are reduced, as shown in Fig.~\ref{img:filters}. 

After data smoothing, a prominence-based peak detection algorithm is implemented to identify signals, as simple threshold-based methods proved inadequate due to noise and overlapping signals. This approach, which considers both prominence and minimum peak distance, significantly enhances peak identification accuracy, as shown in Fig.~\ref{img:filterspeaks}.

Given these exemplary data, the algorithm's performance varies between filters. The precision and recall can be used for verification. 
\ac{tp} denote the number of correctly identified positive instances, \ac{fp} the number of incorrectly identified positive instances, and \ac{fn} the number of missed positive instances. Precision and Recall are defined as follows:
\begin{equation*}
\mathrm{\text{Precision}} = \frac{\mathrm{\text{TP}}}{\mathrm{\text{TP}} + \mathrm{\text{FP}}}
\end{equation*}
\begin{equation*}
\mathrm{\text{Recall}} = \frac{\text{TP}}{\text{TP} + \text{FN}}.
\end{equation*}
Precision measures the proportion of correctly identified positives among all instances predicted as positive, whereas Recall quantifies the proportion of correctly identified positives among all actual positive instances.
The F1-score is defined as follows: 
\begin{equation*}
\text{F1-score} = \frac{2 \cdot \mathrm{\text{Precision}} \cdot \mathrm{\text{Recall}}}{\mathrm{\text{Precision}} + \mathrm{\text{Recall}}} = \frac{2 \cdot \text{TP}}{2 \cdot \text{TP} + \text{FP} + \text{FN}}. 
\end{equation*}
The F1-score provides a balanced harmonic mean of precision and recall, offering a single robust metric to evaluate peak detection performance, particularly when dealing with imbalanced datasets where false positives and false negatives carry comparable costs. 
For the \ac{maf} two additional peaks are detected, but one is incorrectly identified as noise. In contrast, the \ac{kf} is more robust against noise, but a subtle peak is missed due to its smoothing characteristics. Considering multiple recordings with a total of 280 induced ``high'' signals, \autoref{tab:Comparison} shows a comparison of precision, recall and the F1-score for the data with and without preprocessing the data. 
\begin{table}[htb]
     \centering
     \caption{Comparison of Performance of Peak Detection using the Raw Data, \ac{kf} and \ac{maf}}
     \small
     \begin{tabular}{lccc}
     \toprule
          &  Raw Data & \ac{kf} & \ac{maf} \\ 
         \midrule
         Precision & 57.55\,\% & 98.78\,\%  &  91.00\,\% \\
         Recall & 87.05\,\% & 87.14\,\% &  87.05\,\%\\
         F1-score &  69.42\,\% & 92.60\,\% & 94.10\,\%  \\
         \bottomrule
     \end{tabular}
     \label{tab:Comparison}
\end{table} 
 
These findings indicate an significant improvement when applying the filtering techniques in a preprocessing step for the precision. 
However, they also highlight the trade-offs between noise reduction and signal preservation in microbubble-based communication systems, when comparing the results for the two different filtering approaches.
Analysing the recordings confirms as expected that the \ac{kf} is particularly effective for data with a high degree of noise, as it substantially reduces noise while preserving the integrity of the signal. This is evident in Table~\ref{tab:Comparison}, where the F1-score improves from $69.4\,\%$ for raw data to $92.6\,\%$ after \ac{kf} processing, demonstrating its superior performance in noisy conditions. 
Conversely, the \ac{maf} is more suitable for data with lower noise, as it enhances signal clarity through additional smoothing. This is reflected in \autoref{tab:Comparison}, where the F1-score increases from $69.4\,\%$ for raw data to $94.1\,\%$ after \ac{maf} processing.
Overall, the achieved accuracies affirm the viability of microbubble-based communication. By employing appropriate data filtering techniques and implementing robust peak detection algorithms, it is possible to reliably identify ``high'' signals generated by microbubbles.



\section{Implications for Coding and System Design} 
\label{sec:coding}

\noindent To demonstrate the feasibility of information transfer using microbubbles, a time-based \ac{ook} modulation scheme is implemented in the experimental setup. Information is encoded within the duration of each microbubble injection and the subsequent idle period. A binary $1$ is encoded by injecting microbubbles for $0.3\,\text{s}$, while a binary $0$ is represented by a $2.0\,\text{s}$ idle period without injection. 
The total symbol duration is thus $T_\text{sym} = 2.3\,\text{s}$, yielding a raw data rate of $$R_\text{b} = 1/T_\text{sym} \approx 0.43\,\text{bit/s}.$$ The ``off'' phase serves channel stabilization and bubble clearance, introducing a relative time overhead of $$\text{O}_{\text{time}}=\frac{\text{T}_{\text{off}}}{\text{T}_{\text{sym}}}=\frac{2.0\,\text{s}}{2.3\,\text{s}} \approx 87\,\%.$$ This asymmetry ensures distinguishability of successive injections while minimizing inter-symbol interference.

\subsection{Data Rate under Uniform Distribution}
Under uniform symbol distribution (equal probability of $0$ and $1$), the expected average bit duration is
$$\text{T}_{\text{avg}}= \frac{0.3\text{s} + 2.0\text{s}}{2} = 1.15\,\text{s}$$
resulting in an effective bitrate of  
$$ \text{R}_\text{b} = \frac{1}{\text{T}_{\text{avg}}} \approx 0.87 \, \text{bit/s}.$$
The active injection time constitutes only
$$ \eta = \frac{0.3\text{s}}{1.15\text{s}} \approx 26\,\%$$ of $\text{T}_{\text{avg}}$, with the remainder being idle overhead. This trade-off reflects physical constraints of microbubble dynamics but enables reliable ``high'' signal separation, as seen in \autoref{sec:results}. 

\subsection{Reliability Metrics}


Building on the ``high'' signals detection performance from \autoref{sec:results} (F1 scores: KF 92.6\,\%, MAF 94.1\,\%, raw data 69.42\,\%), this section introduces the communication metric \ac{ber}.
The \ac{ber} is defined as:  
$$\ac{ber} = \frac{\text{Number of Bit Errors}}{\text{Total Number of Bits sent}}.$$
In this work with using \ac{ook}, each detected ``high'' signal represents a bit with the value 1, so \ac{ber} is computed as 
$$\ac{ber} = \frac{\text{FP} + \text{FN}}{\text{Peaks}},$$ 
where Peaks denote the total number of ``high'' signals. 
The complementary \ac{bsr} follows as $\ac{bsr} = 1 - \ac{ber}$. \autoref{tab:Comparison2} compares both metrics across filtering approaches.

\begin{table}[htb]
     \centering
     \caption{Comparison of the bit error rate and the bit success rate using the Raw Data, \ac{kf} and \ac{maf}}
     \small
     \begin{tabular}{lccc}
     \toprule
          &  Raw Data & \ac{kf} & \ac{maf} \\ 
         \midrule
         BER & 76.79\,\% & 13.93\,\%  &  11.79\,\% \\
         BSR & 23.21\,\% & 86.07\,\% &  88.21\,\%\\
         \bottomrule
     \end{tabular}
     \label{tab:Comparison2}
\end{table} 
Reprocessing substantially reduces \ac{ber} from $76.79\,\%$ (raw data) to $13.93\,\%$ (\ac{kf}) and $11.79\,\%$ (\ac{maf}), achieving \ac{bsr} values above $86\,\%$. 
These results confirm robust ``1''-bit detection despite noise and recirculation effects.




\subsection{Coding Strategies for Sparse Channels} 
With achieved reliability (\ac{bsr} $>86\,\%$ corresponding to \ac{ber} values of $11.79$--$13.93\,\%$ after filtering), deployment of higher-layer protocols becomes feasible, while the low bitrate ($0.43$--$0.87\,\text{bit/s}$) and temporal correlation from recirculating bubbles necessitate coding schemes adapted to sparse, asymmetric channels. 
Network coding with short block lengths could exploit detected ``high'' signals for redundancy, compensating $1\to0$ errors (dominant in FN). Future work will investigate rateless codes tailored to this channel memory, targeting IoBNT control messages.

\section{Discussion and Outlook}
\label{sec:discussion}

\noindent The experimental testbed described in this work offers a controlled environment that is crucial for evaluating the potential of microbubbles as data transmitters in ultrasound-based communication systems. Utilising the ultrasound recordings that have undergone preprocessing with the \ac{kf} and \ac{maf} algorithms, the efficacy of signal processing and detection is demonstrated.

Despite these advancements, several challenges persist. Overlapping, recirculating, or persistent microbubbles introduce complexities, particularly over extended periods, where residual bubbles contribute to baseline noise and hinder peak detection. Additionally, limitations in the current experimental setup, such as the relatively large tube size and suboptimal injection method, result in interference and inconsistencies that affect data accuracy. These issues highlight the need for improvements in both experimental design and data processing techniques to optimize performance in dynamic environments.

Future work will focus on improving the accuracy and reliability of microbubble detection. One priority is the optimization of filtering parameters to enhance the sensitivity and specificity of peak detection, particularly in complex and dynamic fluid systems. Additionally, integrating machine learning techniques, such as the Autoregressive Integrated Moving Average Model or other anomaly detection algorithms, offers the potential for adaptive and real-time filtering tailored to the unique characteristics of microbubble flow.

Another important direction involves validating the findings under diverse experimental conditions. This includes testing different setups, flow rates, and environmental parameters to ensure that the proposed methods are robust and generalizable across a wide range of scenarios. At the same time, improvements to the experimental setup itself are crucial. Planned modifications include reducing the tube size for greater control, minimizing interference to achieve cleaner signal acquisition, and refining the injection method to enhance the consistency and precision of bubble generation.

Channel encodings tailored to this sparse, asymmetric medium are investigated, including short block polar codes for errors with predominantly false-negative results and rateless fountain codes that utilize temporal correlations from bubble feedback.

%
%
%
%
\section{Conclusion}
\label{sec:conlusion}

\noindent This work demonstrates the feasibility of microbubbles as information carriers in \acl{mc} systems through an experimental testbed. Utilizing the GAMPT BCF300 Doppler ultrasound sensor and clinically-approved SonoVue microbubbles ($\sim$ 2.5\,\textnormal{\textmu{}m}) a reliable detection despite challenging conditions is demonstrated. 
These advancements position microbubble-based \acl{mc} as a foundational technology for Internet of Bio-Nano Things (IoBNT) networks, enabling hybrid diagnostic and communication implants integrated with 6G biomedical systems. 

Future work will focus on improving the accuracy and reliability of microbubble detection by optimizing filtering parameters and exploring adaptive approaches such as machine learning–based anomaly detection. 

%
%
%
%
\section*{Acknowledgment}
This work has been supported by the Federal Ministry of Research, Technology and Space
of the Federal Republic of Germany (F\"{o}rderkennzeichen 16KIS1990,  
IoBNT) in Cooperation with (F\"{o}rderkennzeichen 16KIS2402K, Open6GHub+). The authors alone are responsible for the
content of the paper. 


%
%
%
%

\bibliography{references}  

\end{document}